# Clickbait Detection using Multiple Categorization Techniques

Abinash Pujahari[1] and Dilip Singh Sisodia[2]

[1,2]Department of Computer Science and Engineering
National Institute of Technology Raipur,India

**Abstract.** Clickbaits are online articles with deliberately designed misleading titles for luring more and more readers to open the intended web page. Clickbaits are used to tempted visitors to click on a particular link either to monetize the landing page or to spread the false news for sensationalization. The presence of clickbaits on any news aggregator portal may lead to unpleasant experience to readers. Automatic detection of clickbait headlines from news headlines has been a challenging issue for the machine learning community. A lot of methods have been proposed for preventing clickbait articles in recent past. However, the recent techniques available in detecting clickbaits are not much robust. This paper proposes a hybrid categorization technique for separating clickbait and non-clickbait articles by integrating different features, sentence structure, and clustering. During preliminary categorization, the headlines are separated using eleven features. After that, the headlines are recategorized using sentence formality, syntactic similarity measures. In the last phase, the headlines are again recategorized by applying clustering using word vector similarity based on t-Stochastic Neighbourhood Embedding (t-SNE) approach. After categorization of these headlines, machine learning models are applied to the data set to evaluate machine learning algorithms. The obtained experimental results indicate the proposed hybrid model is more robust, reliable and efficient than any individual categorization techniques for the real-world dataset we used.

## 1. Introduction

The use of online news media has increased these days rapidly due to the excessive use of the internet. These are very useful for the users in gathering knowledge and information at any time, however, sometimes these websites create frustration and waste the time of users by providing altered content than the news headlines. These days most of the websites are using unwanted 'Advertisements/News' kind of things to make money out of it. One of the examples of this is the usage of "Clickbait" [1], [2] headlines. These are the headlines which appear on news websites that attract users and forces them to click on those headlines so that the website can earn money from users' clicks [3][4]. The information present in these headlines creates suspense and can tease users by containing exaggerate information than actual content. The main aim of these (clickbait) headlines is to lure users to click on the headlines. Finally, it causes a lot of frustration for the users. Some of the common clickbait headline examples are given in Table *1*.

A lot of work has been done to combat the clickbait titles. Some tools are available in different leading media sites which automatically blocks such articles. Bauhaus Universitat Weimar Organized a clickbait challenge[1] to detect clickbait by providing their data sets, which draws a lot of attraction in this domain of research. But the problem is that the structure of these headlines is quite similar to non-clickbait headlines which cause a problem. This paper aims to provide an efficient method to categorize the clickbait and non-clickbait articles using semantic analysis and validate using machine learning classification methods. The main contributions of this paper are as follows:

- Categorizing titles to ‘clickbait’ and ‘non-clickbait’ using document formality measures like F-Score[5] and Coh-Metrix[6], because most of the clickbait articles have poor sentence structures.
- Used word to vector scheme for finding the similarity among the texts among both the categories (clickbait and non-clickbait) for proper categorization.
- Build and validate a hybrid model by using above categorization techniques for detection of clickbait headlines by using different machine learning algorithms.

[1] www.clickbait-challenge.org

Table 1. Example of Clickbait Headlines

| Headlines | Description |
|---|---|
| “Man tries to hug a wild lion; you won’t believe what happens next.” | These kinds of headlines seem to be shocking, amazing and unbelievable which generates curiosity among users. |
| “Remember the girl played the role of ‘Nikita’ in the movie ‘Koi Mil Gaya’?” This is how she looks now! Absolutely hot! | These kinds of ‘celebrity gossip’ headlines are teasing contents which force users to click on the headlines. |
| “Only the people with an IQ above 160 can solve these questions. Are you one of them? Click to find out…” | These headlines make a challenge to our IQ, which creates anxiety to explore, but the content may be different |

This paper is organized as follows. The next section contains some related works/methodologies available for detecting clickbait headlines. Section 3 contains the proposed methodology used in this paper to detect clickbait article. Section 4 contains the result and discussion of the performance of proposed methodology along with the performance of different machine learning algorithms for detecting clickbait headlines. Finally, the conclusion section contains the general outcome of this paper along with the future scope of research in this field.

## 2. Related Works

Some of the related works in the domain of clickbait detection is described in details in this section as well as their limitation and possible extensions.

Chakraborty et al. [7] proposed a method for detecting clickbait articles and also built a browser add-on for detecting clickbaits. They have collected non-clickbait articles ‘Wikinews’, and for clickbait article, they followed several domains like (BuzzFeed, Viral Nova, etc.). They carried out linguistic analysis on the dataset collected using the ‘Stanford CoreNLP’ [8] tool. They primarily focused on the Sentence structure (i.e., Length of the headline/words, hyperbolic words, internet slang, common phrases, determiners) to categorize the clickbait and non-clickbait headlines. They used four feature selection techniques: Sentence Structure, Word Patterns, Clickbait Language, N-gram Features. Finally, they compared the classification of articles using three machine learning algorithms: SVM, Decision Tree, Random Forest.

Another work was done by Biyani et al. [9] to detect clickbait, in which they first categorized the type of clickbait headlines into eight categories (Exaggeration, Teasing, Inflammatory, Formatting, Graphic, Bait-and-switch, Ambiguous, Wrong). According to them, these eight are the most common categories of clickbait headlines. They have used document informality measure and reading difficulty of the text to determine whether a headline is a 'Clickbait' or not. They applied 'Gradient Boosted Decision Trees' classification algorithm to the data set using four features (Similarity, URL, Content and Informality and Forward Reference) and test them individually using test data. They have also made the comparison of the density of clickbait and non-clickbait headlines. The performance is acceptable in this paper for most of the data set, but the individual features used is not enough to get the desired outcome.

Rony et al. [10] extended the prevention of 'Clickbait' to a social media platform. They have used the Skip-Gram model [11] to use word embeddings, which is further used to find out the similarity between the texts. They have made the comparison of their pre-trained vectors with the Google news dataset. They compared the results with earlier findings in the field of clickbait detection with pre-trained vectors and without pre-trained vectors. They have also categorized the percentage of different media in terms of clickbait and non-clickbait. They have used the headline-body similarity mainly for categorizing the clickbait article.

The main problem of clickbait detection is the categorization of headlines. Because earlier research suggests that sentences having the similar kind of structure and context can fall into either of the categories (Clickbait and Non-clickbait). The next section describes the proposed method for detecting clickbait headlines.

# 3. Methods

To improve the quality of the clickbait detection techniques discussed earlier in section 2, this paper used the model as represented in Fig. 1 for clickbait prevention. The dataset is obtained from [7] which contains news headlines of both kinds (clickbait and non-clickbait). The data set is thoroughly evaluated using the proposed method. After categorization of the data into two classes, they are evaluated using the machine learning algorithms, which are described in later sections.

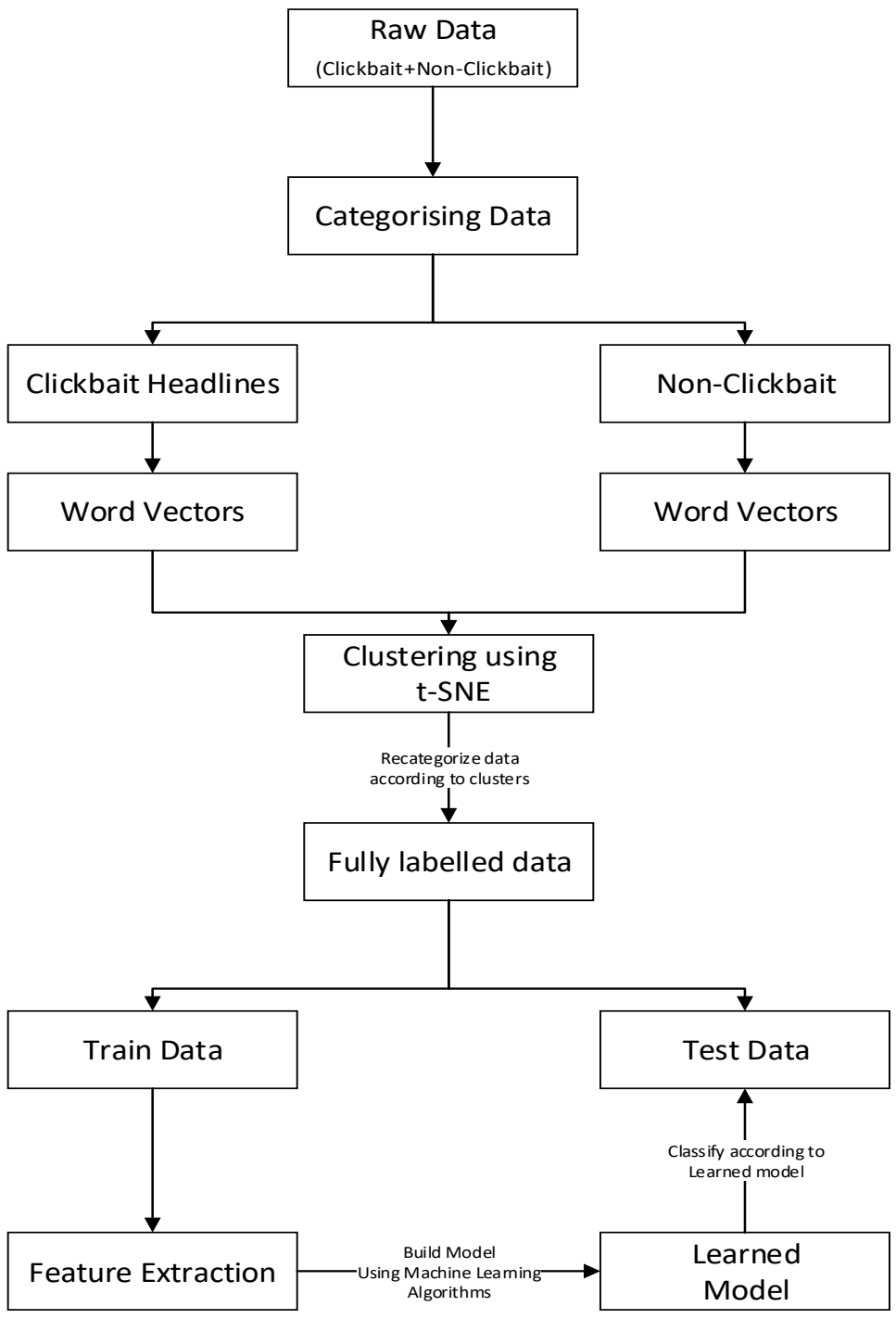

Figure 1 Proposed Model for Clickbait Classification

## 3.1 Categorisation of Headlines

The proposed model of this paper extends the categorization of clickbait headlines by Biyani et al. [9], by adding more constraints in categorization. The proposed constraints are given in Table 2. This helps in identifying the clickbait and non-clickbait articles more precisely. Further, we would like to mention that, the categories mentioned in Table 2 are not the subsets of any of the categories mentioned by Biyani et al. These are the constraints we applied in addition to the eight constraints available in [9]. The purpose of this addition is to classify a headline to which category it belongs more correctly. After adding these constraints, we are getting almost 10% of the data from the dataset were recategorized, which is analyzed later in this paper, which improves the performance of categorization. Some of the headlines are falling to more than one category, which is trivial. The major category out of the three categories from Table. 2 is 'Incomplete' because it always creates suspense among users and forcing them to click on the headline. Out of these three, the headline cloning is very difficult to detect. Because the normal headlines and the 'clickbait' are almost the same. To detect this kind of clickbaits, we parsed the body of the landing document and compared the keywords with the headlines in order to determine whether it is a clickbait or not.

Table 2 Categories of Clickbait Headlines and their Examples

| Category | Definition | Example |
|---|---|---|
| Incomplete | The title is incomplete in conveying the message | "Click here, and you will get…" |
| Headline Cloning | Copying of actual headline for different content. | Headlines have a different structure but same text as normal headlines. |
| URL Redirection | The headlines lands on a false page than promised | Invalid URLs having false domain information (e.g. http://xyz.by) |

## 3.2 Sentence Formality and Structure

To find out the sentence formality of the headlines we have used two features. The first one is F-Score [5] which is also used by (Biyani et al. [9]) and the second is the Coh-Matrix. F-score, which is calculated using Eq. (1), has a value of 0 to 100%, i.e., the higher the value, the more formal the language is. The threshold percentage for our experiment is 60% as F-Score value. We have taken the intersection part between these two schemes to determine the well-formed headlines. Because the 'Clickbait' headlines are poorly formed. To find out the terms listed in Eq. 1 and analysis of word in the text, we have used the Stanford CoreNLP [12] language tool.

$$F - Score = (noun\ freq. + adjective\ freq. + preposition\ freq. + article\ freq - pronoun\ freq. - verb\ freq - adverb\ freq - interjection\ freq + 100)/2 \quad (1)$$

Coh-metrix[6] is used to find the ambiguous words, syntactic complexity, word ratio, readability, co-reference cohesion etc. After using these two measures, two refined sets of headlines are created (clickbait and non-clickbait). Coh-Metrix[2] is an online tool for accessing the features mentioned above from a piece of document. The headlines we collected is evaluated online using this tool to separate clickbait headlines and normal headlines. This feature is used along with F-Score because it accesses the document up to multiple levels.

[2] Coh-Metrix tool is available here: http://cohmetrix.com/

This paper also uses the Flesh-Kincaid grade level [13] which is used for readability test for different text reading applications. In general practice, the 'clickbaits' are more difficult to be read, due to their structure, whereas the normal headlines are easier to be read. The formula for the Flesch reading-ease score (FRES) [13] test is given in equation 2.

$$Score = 206.835 - 1.015\left(\frac{total\ words}{total\ sentences}\right) - 84.6\left(\frac{total\ syllables}{total\ words}\right) \quad (2)$$

The value of the Score is in between 0 to 100. Again, the higher the value, easier is the readability of the text. So, the headlines having high scores are generally normal headlines, and low score headlines can be 'clickbait'. The threshold value for the normal headlines for Eq. (2) is above 60 in our experiment to distinguish between clickbait and non-clickbait.

### 3.3 Recategorization of Headlines using Clustering

By using the processes mentioned above, the raw data obtained is divided into two categories. But still, the data is noisy since the volume of the data is very large. The next step of this work is to recategorize the headlines. For this, we use the word representation in vector space [14] by using the word to vector scheme. After converting the words to word vectors, we have used the t-SNE [15] algorithm to create two clusters where similar words are grouped together. This is a dimension reduction approach to categorize high dimensional data. The t-sne method is proposed by Van Der Maaten et al., which is an extension to Stochastic Neighbourhood Embedding(SNE), proposed by Hinton et al. [16]. SNE method uses probability distribution over pairs of high-dimensional objects such that, similar objects have a high probability of being grouped, while dissimilar objects have a very small probability of being grouped. For calculating the similarity between two objects, they calculated the conditional probability using Eq. (3) [16], where $P(x(j)|x(i))$ is the conditional probabilty that $x(i)$ and $x(j)$are treated as neighbours.

$$P(x(j)|x(i)) = \frac{\exp\left(-\|x(i) - x(j)\|^2/2\sigma_i^2\right)}{\sum_{k\neq i} exp\left(-\|x(i) - x(k)\|^2/2\sigma_i^2\right)} \quad (3)$$

Here $\sigma_i$ is known as Gaussian variance which is centered on the object $x(i)$. t-SNE is differena t method for embedding of objects by overcoming the 'Crowding problem' and uses t-distribution rather than Gaussian distribution as proposed Van Der Maaten et al. Before applying these procedures, we have converted the data into word vectors by using word vecthe tor embedding technique. The results of the neighbourhood embedding are discussed in the next section.

## 4. Experimental Results and Discussion

The data set is obtained from [7], which contains the normal headlines as well as clickbait headlines. First, the data set is categorized between normal text and texts having marks like (punctuation, exclamation, question) because these are often present in clickbait headlines. After doing this, the documents are categorized into clickbait and non-clickbait by using the features discussed in section 3.1 by comparing the features of the headlines. The clickbait articles are grouped into total 11 categories. The number of headlines falling into these categories is given in Table 3. One headline can fall into more than one category due to the sentence structure and text format. Hence, the total number of headlines classified into each category exceeds the total no. of clickbait headlines. The categories from serial number 1 to 8 were taken from [9], and the rest are proposed in this paper. The precision of each category is also mentioned in Table 3. The headlines not falling into these categories are grouped as non-clickbait headlines during this phase of categorization.

Table 3 Precision of different Clickbait Categories

| SI No. | Category | No. of examples classified as clickbait | Precision |
|---|---|---|---|
| 1 | Ambiguous | 645 | 47.81% |
| 2 | Exaggeration | 4954 | 45.86% |
| 3 | Inflammatory | 1023 | 52.34% |
| 4 | Bait-and-switch | 536 | 65.82% |
| 5 | Teasing | 5278 | 59.81% |
| 6 | Formatting | 789 | 49.63% |
| 7 | Wrong | 152 | 41.23% |
| 8 | Graphic | 365 | 40.78% |
| 9 | Incomplete | 678 | 48.64% |
| 10 | Headline Cloning | 831 | 51.56% |
| 11 | URL Redirection | 1345 | 53.96% |

After categorizing the headlines into clickbait and non-clickbait using the features mentioned in Table 3, we refined the grouping using the document formality measures as discussed in Section 3.2 using F-Score and Score measures using Eq. (1) and Eq. (2) respectively.

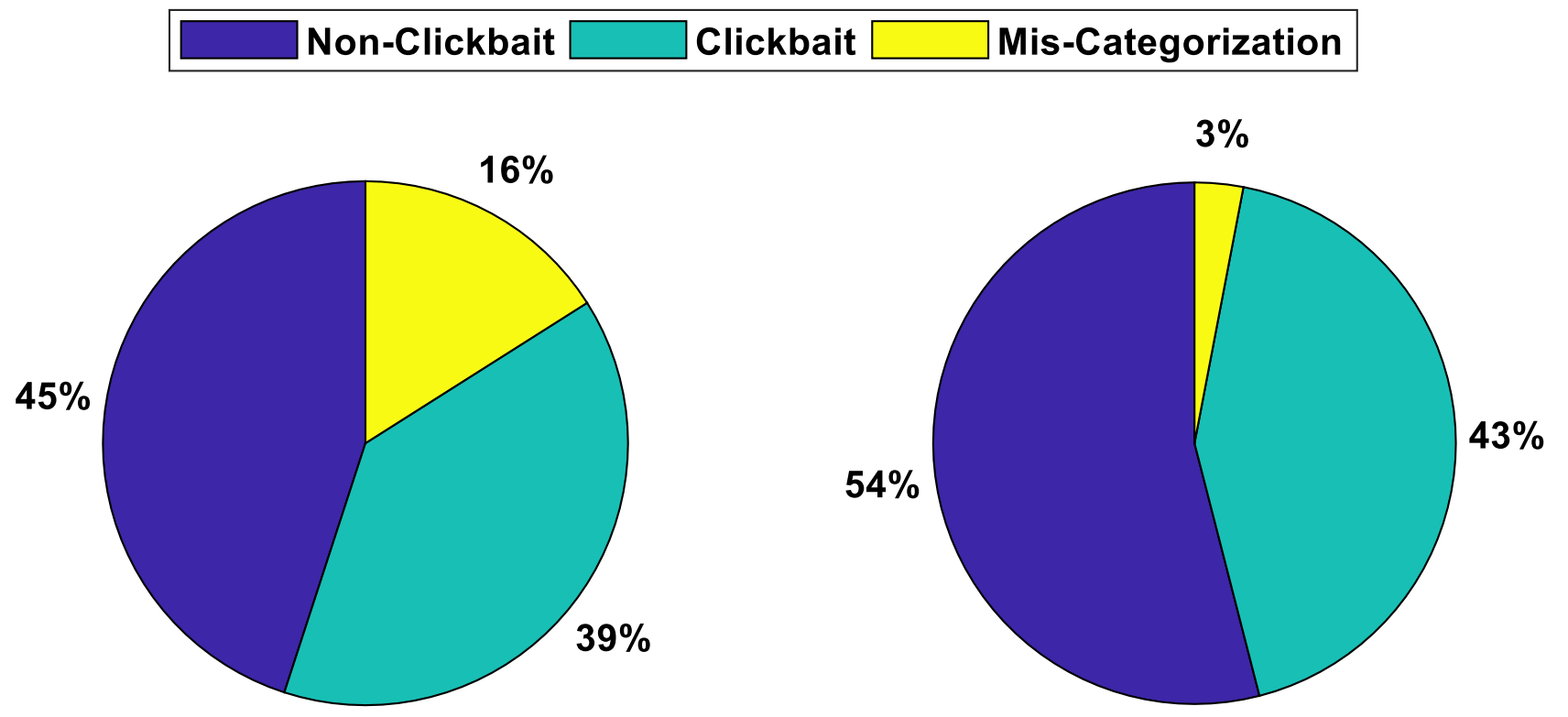

Figure 2. Percentage of Clickbait and non-clickbait article after clustering

After grouping the headlines into two categories, word embedding vector is generated for both types of headlines. Then the words are converted to word vectors using word2vector using MatLab Text Analytics[3] toolbox. After converting to vector format, then the data set is applied through clustering using t-distributed stochastic neighborhood embedding technique as discussed in Section 3. We can clearly see from Fig. (2) That, more numbers of examples are grouped into either of the categories

[3] Text Analytics Tool box MATLAB: https://in.mathworks.com/products/text-analytics.html

(i.e., clickbait or non-clickbait). This makes the two classes (clickbait and non-clickbait headlines) more robust and less noisy. Further, it can be seen from the figure that the wrong categorization of headlines is less after using word vector clustering. A wrong categorization is that if a headline is actually clickbait and it is categorized as non-clickbait and vice-versa. The two different clusters (clickbait and non-clickbait headlines) are represented in text scattered plots in Fig. (3). From Fig. (3), it can be seen that there is very less noise in the data after categorization and word vector clustering. This data is now ready for classification and model learning procedures.

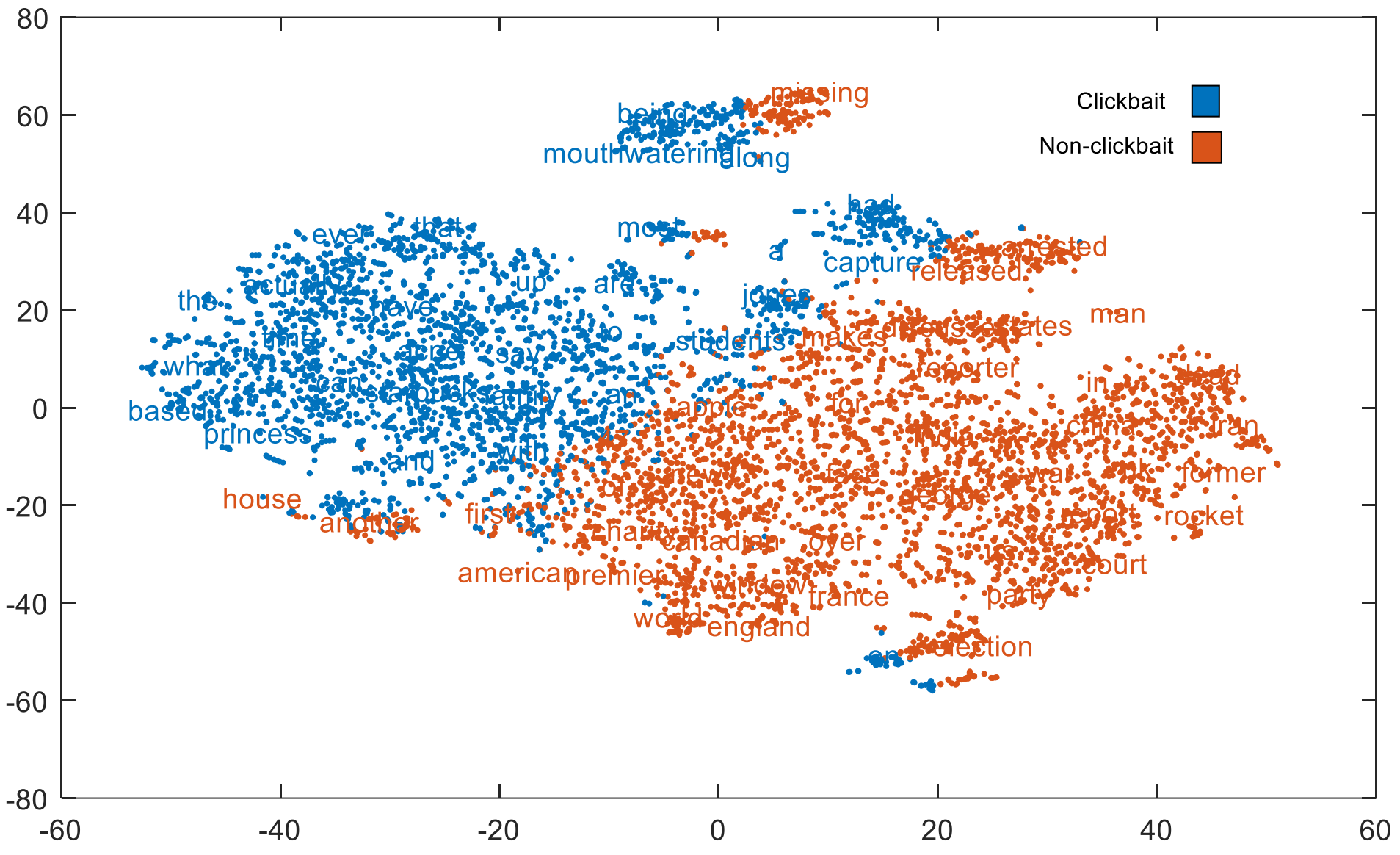

Figure 3. Clustering of headlines using t-SNE

## 4.1 Classification of Dataset using Learning Algorithms

After going through the rigorous categorization of headlines, now a model is built using learning algorithm so that it can predict the unseen examples in the future. We have used the word vectors as our features for training purpose. The classification algorithms used in this paper are ‘Support Vector Machine’[17], ‘Decision Tree (C4.5)’ [18] and ‘Random Forest’ [19]. We have used the 10-fold cross-validation scheme to test the efficiency of the classifiers. The results were first evaluated individually using different features and then evaluated after integrating all the features. The evaluation criteria used for the classifiers are Accuracy, Precision, Recall. The ROC curves for the respective classifiers are also generated to determine the relationship between true positive rates and the false positive rates.

Table 4. Performance of different classifiers with respect to features

| | **Decision Tree** | | | **SVM** | | | **Random Forest** | | |
|---|---|---|---|---|---|---|---|---|---|
| **Features** | **Accuracy** | **Precision** | **Recall** | **Accuracy** | **Precision** | **Recall** | **Accuracy** | **Precision** | **Recall** |
| Based on Categories(C) | 0.86 | 0.89 | 0.87 | 0.92 | 0.94 | 0.90 | 0.88 | 0.92 | 0.89 |
| Based on Structures(S) | 0.84 | 0.89 | 0.81 | 0.90 | 0.92 | 0.89 | 0.89 | 0.91 | 0.88 |
| C + Word Vector Clustering | 0.91 | 0.92 | 0.89 | 0.95 | 0.95 | 0.95 | 0.93 | 0.94 | 0.91 |
| S + Word Vector Clustering | 0.90 | 0.91 | 0.88 | 0.93 | 0.84 | 0.93 | 0.91 | 0.91 | 0.90 |
| All Features | 0.92 | 0.93 | 0.92 | 0.97 | 0.97 | 0.96 | 0.94 | 0.94 | 0.93 |

The performance of the classifiers used in this paper is given in Table 4. It can be seen clearly from the table that, the individual features are not performing consistently, but when integrating all the features, the results are quite acceptable for the data used for clickbait headline detection. The results of the individual features are less as compared due to the inconsistencies in the data obtained from different online media. Because the same headlines can have different structures, different semantics, and different sentence formation. Hence, by integrating all the features we are getting good results. The confusion matrices of the respective classifiers are shown in Fig. (4). The Receiver Operating Characteristic (ROC) curves for different classifiers are also shown in Fig. (5). We are getting AUC values, 0.94, 0.96 and 0.99 for Decision Tree, Random Forest and SVM respectively which are quite competent to earlier techniques mentioned in this paper for the data set we have used. Figure 6 represents the AUC graphs for the respective classifiers. Comparing among the classifiers, SVM is performing better than other two classifiers in clickbait headline detection, which can be clearly seen from the confusion matrices as well as ROC curves. SVM performs well on text analytics and classification using the linear kernel because text data contains a lot of features and they are linearly separable in most of the times. Since the dataset we have used is a two-class problem, SVM is performing better than other classifiers used in the experiment.

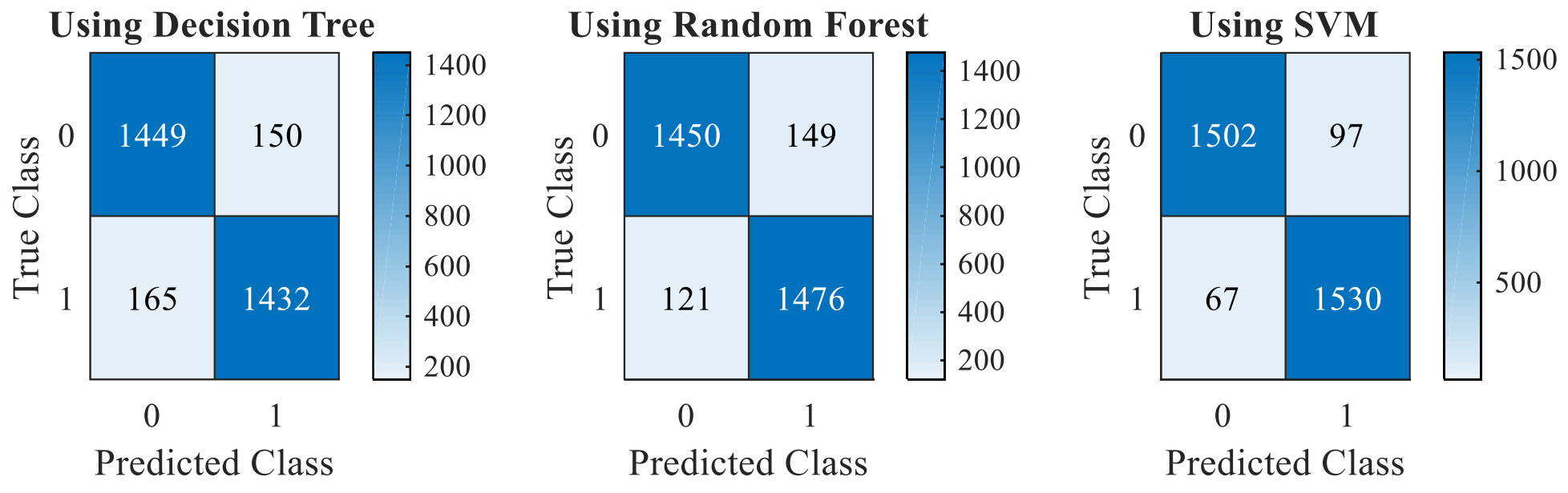

Figure. 4 Confusion Matrices along with Heatmap for different classifiers

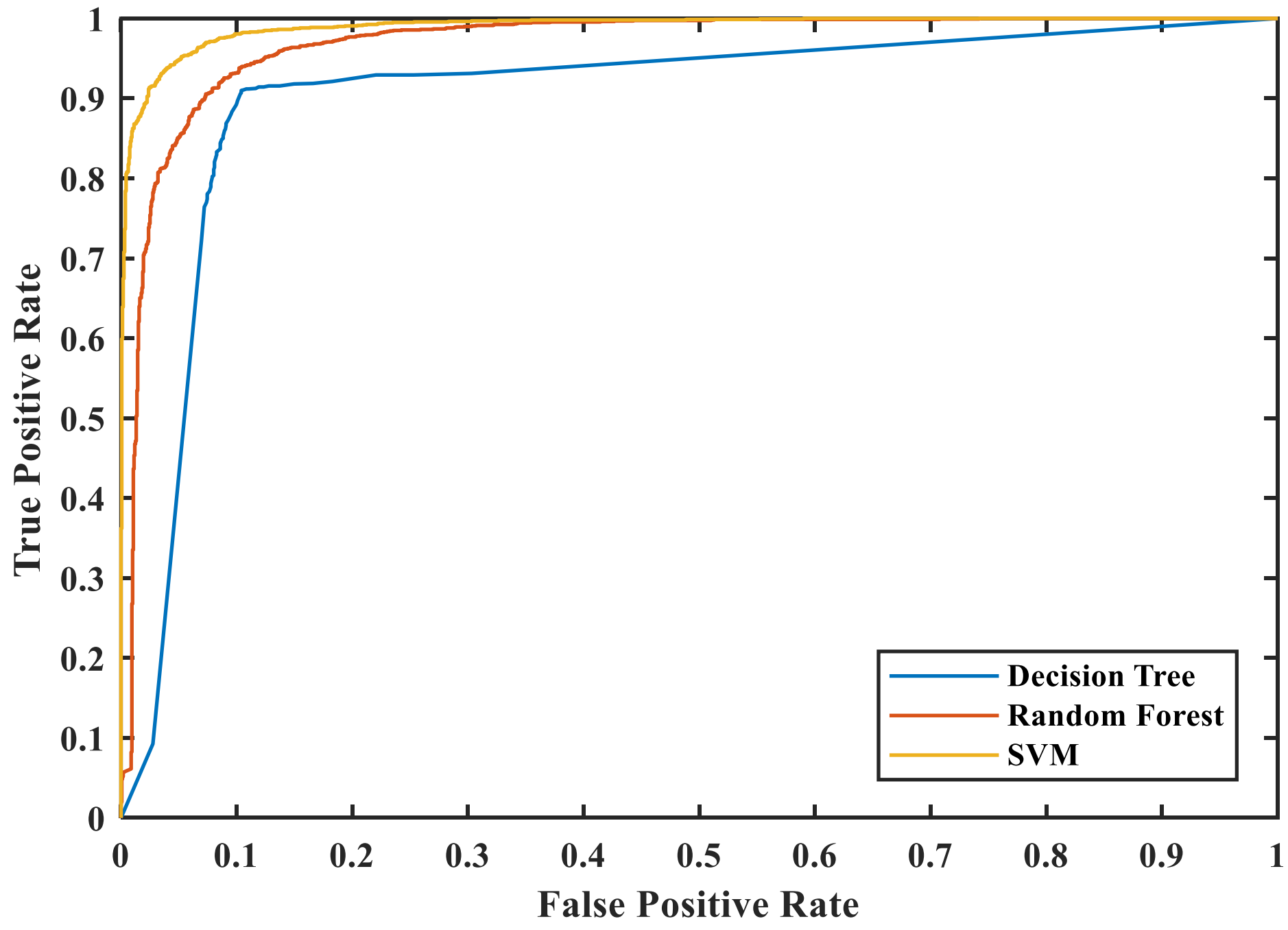

Figure 5. ROC Curves for different Classifiers

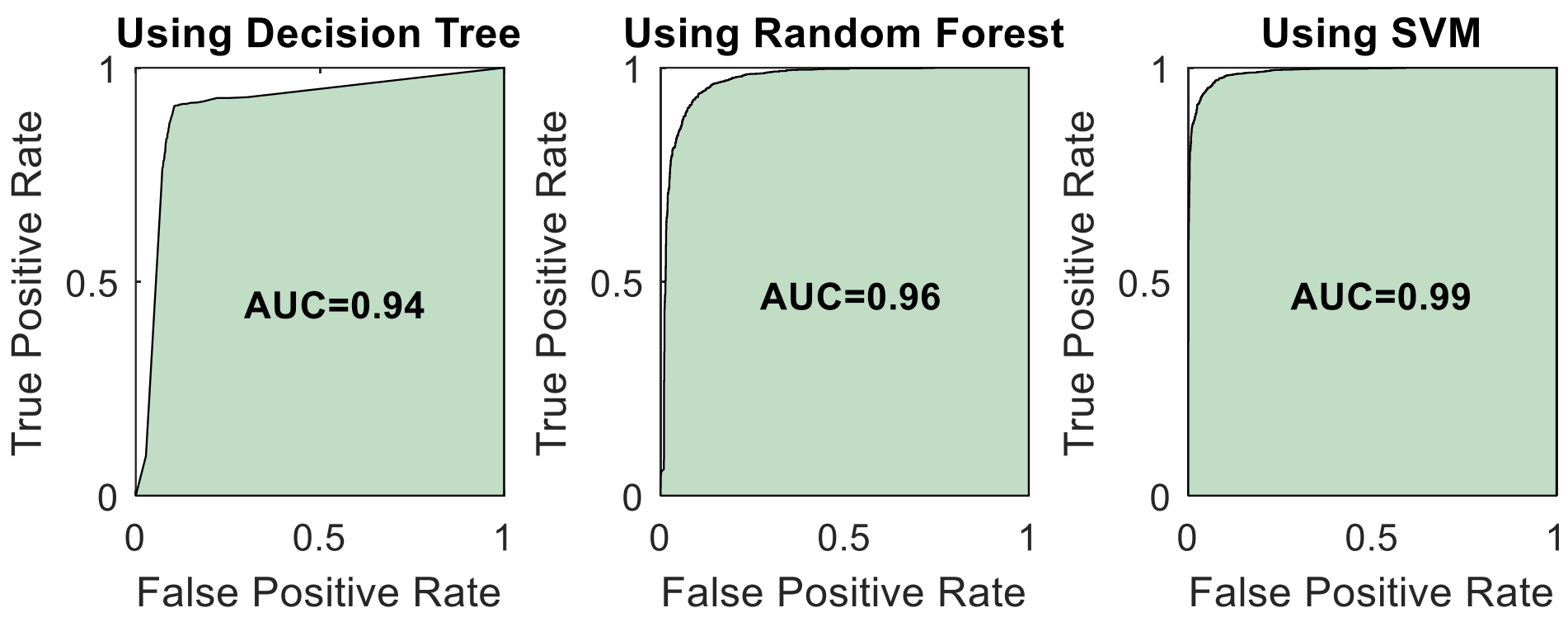

Figure 6. AUC values for different classifiers

## 4.2 Reliability Test of the Model for Detecting Clickbait

Alexandru et al. in their paper [20] proposed the model for predicting good probabilities for supervised learning. They also proposed two calibration methods for correcting the distortions generated by the bias for different classifiers. Using the same procedure, we examined the relationship between the predictions made by the learning mentioned above with true posterior probabilities Reliability graph allows us to check if the predicted probabilities of a binary classifier are well calibrated. Since our classification problem is binary, we can test the classifiers using reliability graph. In reliability graph, the curve should be as close as possible to the diagonal/identity. The respective reliability graphs are represented in Fig. (7) For different classifiers. As we can see from the graph, the curve is closest to the diagonal in case of SVM classifiers as compared to other two classifiers.

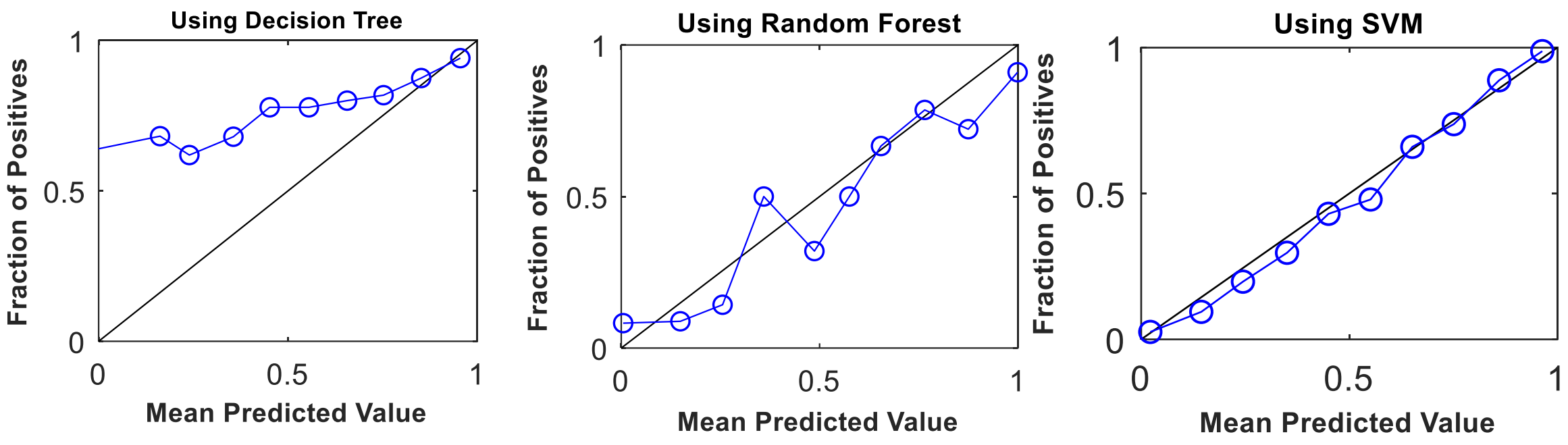

Figure 7. Reliability Graph of Different Classifiers by integrating all the features for detecting clickbait

# 5. Conclusion

Clickbait headlines have become a major issue in online news media. Hence the automated prevention is necessary. We can conclude from the experimental results section that, only one categorization technique is not efficient enough to combat clickbait articles. The results are quite acceptable by integrating more than one feature. Further, the websites are changing their strategies with respect to time also. Hence, the manufacturer will also find out the bypass the categorization of clickbait articles. Hence a robust detection should be build that can predict future changes in the detection procedure by taking the time domain features. Also, the websites are using graphical images as clickbait headlines, which cannot be detected using text processing. So, by applying some image processing and pattern recognition schemes, we can detect the same after extracting the sentences. These can be the future scope of research in this domain.